\documentclass[aps,prl,twocolumn,superscriptaddress]{revtex4-1}
\usepackage{%epsf,
amsfonts,amssymb,epsfig,amsmath,graphics,subfig}
\usepackage{color}
\usepackage{hyperref,graphicx}
\usepackage{cancel}

\newcommand{\nn}{\notag \\}

\begin{document}

\title{Navier-Stokes Equations on Black Hole Horizons\\ and DC Thermoelectric Conductivity}

\author{Aristomenis Donos}
\affiliation{Centre for Particle Theory and Department of Mathematical Sciences, Durham University, Durham, DH1 3LE, U.K.}
  \author{Jerome P. Gauntlett}
\affiliation{Blackett Laboratory,
  Imperial College, London, SW7 2AZ, U.K.}
\begin{abstract}
Within the context of the AdS/CFT correspondence
we show that the DC thermoelectric conductivity can be obtained by solving the linearised, time-independent and forced Navier-Stokes equations on the black hole horizon for an incompressible and charged fluid.
\end{abstract}
\maketitle

\setcounter{equation}{0}

\section{Introduction}
A striking feature of the AdS/CFT correspondence is that some fundamental properties of the dual conformal field theory (CFT) are captured by the geometry of the black hole horizon. The temperature of the CFT is equal to the Hawking temperature of the black hole which is determined by the surface gravity of the black hole. Similarly, the entropy of the CFT is the Bekenstein-Hawking entropy which is given by one quarter of the area of the black hole event horizon. Although not universal\footnote{In general, the result $\eta=s/4\pi$ will not be valid for translationally invariant anisotropic black holes nor for holographic lattices.}, it is also interesting that for a sub-class of holographic black holes the shear viscosity, $\eta$, is also captured by the area of the black hole horizon via $\eta=s/4\pi$, where $s$ is the entropy density \cite{Policastro:2001yc,Kovtun:2004de}.

Here we will argue that, remarkably, the DC thermal conductivity and, more generally, the thermoelectric conductivity of the 
dual field theory, are universally captured by physics at the black hole horizon. Specifically, one needs to solve linearised, time independent and forced Navier-Stokes equations for an incompressible charged fluid on the curved horizon. 

The thermal conductivity, a property relevant for all dual field theories, determines the heat current, $\bar Q^i$, that is produced after applying a temperature gradient, $\zeta^i=-\partial_iT/T$, at the level of linear response. If the dual field theory
has a global $U(1)$ symmetry, then with the additional application of an electric field $E^i$, the heat current and electric current, $\bar J^i$, that are produced define the thermoelectric conductivities via:
\begin{align}\label{bigform}
\left(
\begin{array}{c}
\bar J^i\\\bar Q^i
\end{array}
\right)=
\left(\begin{array}{cc}
\sigma^{ij} &T \alpha^{ij} \\
T\bar\alpha^{ij} &T\bar\kappa^{ij}  \\
\end{array}\right)
\left(
\begin{array}{c}
E_j\\\zeta_j
\end{array}
\right)\,.
\end{align}
It is important to emphasise that $\bar Q^i, J^i$ are the total current fluxes, 
defined later.
In seeking applications of the AdS/CFT correspondence to real materials, the DC conductivities are important observables to study. Our results can be used to determine whether or not the dual field theory is a conductor or an insulator and, at a more refined level, the temperature dependence of the conductivities, including the appearance of any scaling laws.

From a theoretical point of view the DC conductivities are somewhat subtle to study,
however, since they are generically infinite unless there is some mechanism for momentum to dissipate. 
A natural framework to study momentum dissipation is provided by ``holographic lattices" \cite{Horowitz:2012ky}. 
Namley, black hole solutions with asymptotic behaviour at the AdS boundary that are associated with adding sources to the dual CFT which break spatial translations. There has been much interest in these black holes since they can realise
metal-insulator transitions \cite{Donos:2012js,Donos:2013eha} as well as novel incoherent metals \cite{Donos:2012js,Donos:2014uba,Gouteraux:2014hca}.

The holographic lattices depend on the holographic radial direction as well as the spatial directions and hence constructing them generically involves solving PDEs.  As a result most examples that have been studied are
one-dimensional lattices, which break translation invariance in just one of the spatial directions. An important exception is provided by Q-lattices \cite{Donos:2013eha} (and similar constructions \cite{Andrade:2013gsa}) whose matter content can be used to break translational invariance periodically in all spatial dimensions while the metric remains translationally invariant.

A method for calculating the DC thermoelectric conductivity for Q-lattices and one-dimensional lattices
was presented in 
\cite{Donos:2014uba,Donos:2014cya,Donos:2014yya}. For these lattices the final result was expressed explicitly
in terms of the black hole solution at the horizon.
This substantially extended a similar result found for the DC electric conductivity at zero charge density and with no momentum dissipation \cite{Iqbal:2008by}.
Here we will show that for generic lattices, breaking translations in all
spatial directions, one cannot obtain such an explicit formula for the DC conductivity. However, the DC conductivity can always be obtained by solving Navier-Stokes equations on the black hole horizon. The earlier results can now be viewed 
as special cases in which the fluid equations can be explicitly solved.

An early connection between gravity and fluids is the membrane paradigm \cite{Price:1986yy}. More recently, in the holographic fluid-gravity correspondence \cite{Bhattacharyya:2008jc}, approximate solutions to the gravity equations are obtained by solving relativistic hydrodynamic equations for the boundary theory via a systematic derivative expansion. 
The Navier-Stokes equations arise after taking
a scaling limit and hence they too can be captured
in a dual gravitational description \cite{Bhattacharyya:2008kq,Fouxon:2008tb}. These connections are applicable in a hydrodynamic limit. 
On the other hand in \cite{Bredberg:2011jq} it was shown how solutions of
Navier-Stokes equations on hypersurfaces in Minkowski space give rise to solutions of Einstein's equations. Here, by contrast, solutions of Navier-Stokes equations on the black hole horizon lead to exact transport quantities, given by specific two point correlators, 
in the deformed dual CFT. In obtaining these results we do not take a hydrodynamic limit of the dual field theory (or any other limit) 
and our results apply to arbitrary horizon geometries that
arise as solutions to the equations of motion.
We expect that the time-dependent 
and non-linear generalisation of the fluid equations that we obtain will also play a role in studying 
holographic lattices, for example in a suitable hydrodynamic limit (e.g. see \cite{Hartnoll:2007ih,Blake:2015epa}).
 We emphasise, though, that our results here already show that independently of the strength of translation breaking effects and the temperature, a hydrodynamic description of DC transport is always possible in terms of a specific fluid living on the black hole horizon.

For simplicity, we focus on holographic lattices of $D=4$ Einstein-Maxwell theory; 
the main results extend very simply to $D\ge 4$ as well as the inclusion of other
matter fields. More details appear in \cite{Banks:2015wha}.

\section{Background black holes}
We consider the $D=4$ bulk action\footnote{For simplicity we have set $16\pi G=1$ and also set the cosmological constant to a convenient value.}
\begin{align}\label{eq:bulk_action}
S=\int d^4 x \sqrt{-g}\,\left(R+6-\tfrac{1}{4}\,F^{2}\right)\,,
\end{align}
The unit radius $AdS_4$ vacuum solution is dual to a $d=3$ CFT with a global $U(1)$symmetry.
We will focus on 
the class of electrically charged, static black holes given by
 \begin{align}\label{eq:DC_ansatz}
ds^{2}=-UG\,dt^{2}+\frac{F}{U}\,dr^{2}+ds^2(\Sigma_2)\,, \qquad
A=a_{t}\,dt\,,
\end{align}
where
$ds^2(\Sigma_2)\equiv g_{ij}(r,x)dx^i dx^j$ is a metric on a two-dimensional manifold, $\Sigma_2$,
at fixed $r$. Also, $U=U(r)$, while $G,F$ and $a_t$ are all functions of $(r,x^i)$.

At the $AdS_4$ boundary, as $r\to\infty$, we have
\begin{align}\label{GandF}
&U\to r^2,\qquad F\to 1,\qquad 
G\to \bar G(x),\nn &g_{ij}(r,x)\to r^2 \bar g_{ij}(x),\qquad
a_t(r,x)\to \mu(x)\,.
\end{align}
The spatial dependence of the boundary metric, given by $\bar G(x)$, $\bar g_{ij}(x)$, corresponds to 
a source for the stress tensor of the dual CFT. Similarly, $\mu(x)$ is a spatially dependent chemical potential for the
global $U(1)$ symmetry. An interesting sub-class of
solutions is associated with adding spatially periodic sources to a CFT in flat space. 
In this case the functions are all periodic in the spatial coordinates $x^i$ and we can, in effect, take $\Sigma_2$ to be 
a torus.

The black hole horizon is assumed to be located at $r=0$. 
By considering the Kruskal coordinate $v=t+\frac{\ln r}{4\pi T}+\dots$ we deduce that the
near horizon expansions are given by
\begin{align}\label{nhexpbh}
U\left(r\right)&=r\left(4\pi\,T+U^{(1)}\,r+\dots\right)\,,\nn%+\cdots\ ,\nn
a_{t}(r,x)&=r\left(a^{(0)}_{t}(x)G^{(0)}\left(x\right)+a^{(1)}_{t}\left(x\right)r+\dots\right)\,,\nn
G(r,x)&=G^{(0)}\left(x\right)+G^{(1)}\left(x\right)r+\dots\,,\nn
F(r,x)&=G^{(0)}\left(x\right)+F^{(1)}\left(x\right)r+\dots\,,\nn
g_{ij}(r,x)&=g_{ij}^{(0)}(x)+g_{ij}^{(1)}(x)r+\dots\,,
\end{align}
where the factor of $G^{(0)}$ in the leading term of $a_{t}(r,x)$ has been added so that electric charge density at the horizon is simply $\sqrt{-g}F^{tr}|_H=\sqrt{-g_0}a^{(0)}_{t}$.

\section{Perturbing the black holes}
We will consider a perturbation that provides sources $E,\zeta$ for the electric and heat currents, respectively, that are linear in $t$. 
Specifically,
generalising \cite{Donos:2014uba,Donos:2014cya,Donos:2014yya}, we study
\begin{align}\label{pertone}
\delta (ds^2)&=\delta g_{\mu\nu} dx^\mu dx^\nu-2t GU\zeta _i dt dx^i\,,\nn
\delta A&=\delta a_\mu dx^\mu-t E_i dx^i+t a_t \zeta_i dx^i\,,
\end{align}
with $\delta g_{\mu\nu}$,$\delta a_\mu$ functions of $(r,x^i)$ while  
$E_i=E_i(x)$, $\zeta_i=\zeta_i(x)$ are one-forms on $\Sigma_2$ and we demand that
\begin{align}\label{closed}
d(E_i dx^i)=d(\zeta _i dx^i)=0\,.
\end{align}
This perturbation solves the time dependence of the equations of motion at linear order.

At the $AdS_4$ boundary
we demand that the fall-off of $\delta g_{\mu\nu}$,$\delta a_\mu$ is 
such that the only sources are parametrised by $E,\zeta$. 
At the black hole horizon, as $r\to 0$, regularity implies that we must have
\begin{align}\label{eq:nh_exp}
\delta g_{tt}&=U\left(\delta g^{(0)}_{tt}\left(x\right)+{\cal O}(r) \right),\,
\delta g_{rr}=\frac{1}{U}\,\left( \delta g_{rr}^{(0)}\left(x\right)+{\cal O}(r)\right),\nn
\delta g_{ij}&=\delta g_{ij}^{(0)}\left(x\right)+{\cal O}(r),
\quad
\delta g_{tr}=\delta g_{tr}^{(0)}\left(x\right)+{\cal O}(r)\,,\nn
\delta g_{ti}&=\delta g_{ti}^{(0)}\left(x\right)-\zeta_{i} GU \,\frac{\ln r}{4\pi T}+{\cal O}(r),\nn
\delta g_{ri}&=\frac{1}{U}\,\left( \delta g_{ti}^{(0)}\left(x\right)+{\cal O}(r) \right)\,,\nn
\delta a_{t}&=\delta a_{t}^{(0)}\left(x\right)+{\cal O}(r),\quad
\delta a_{r}=\frac{1}{U}\,\left(\delta a_{t}^{(0)}\left(x\right)+{\cal O}(r)\right)\,,\nn
\delta a_{i}&=\frac{\ln{r}}{4\pi T}(-E_i+a_t\zeta_i)+{\cal O}(r)\,,
\end{align}
with $\delta g_{tt}^{(0)}+\delta g_{rr}^{(0)}-2\,\delta g_{rt}^{(0)}=0$. Note that 
the logarithm terms combine with the terms linear in time in \eqref{pertone}.

\subsection{Electric and Heat Currents}
We define the bulk electric current density as
\begin{align}
J^i=\sqrt{-g}F^{ir}\,.
\end{align}
At the $AdS_4$ boundary we find that $J^i|_\infty$ is the electric current density of the dual field theory.
The gauge equations of motion, $\nabla_\mu F^{\mu\nu}=0$, imply
\begin{align}\label{jandfeoms}
\partial_iJ^i=0\,,\qquad
\partial_rJ^i=\partial_j\left(\sqrt{-g}F^{ji}\right)\,,
\end{align}
and another equation which won't play a further role.

For the heat currents, we want to identify equations of motion involving the metric perturbation 
that have a similar structure to the gauge equations of motion. 
First consider a vector $k$ which satisfies
\begin{align}\label{kcon}
\nabla_\mu k^\mu=0\,,\qquad
\nabla_\mu\nabla^{(\mu}k^{\nu)}=\alpha k^\nu\,,
\end{align}
for some function $\alpha$, which would vanish if $k$ is a Killing vector.
We also write $\varphi={k}^\mu A_\mu$ and $k^\mu F_{\mu\nu}=\partial_\nu \theta +s_\nu$, with $s$ a one-form and $\theta$ an globally defined function.
In the special case that the Lie derivative of $F$ with respect to $k$ vanishes we have $\partial_{[\mu}s_{\nu]}=0$.
We now define the two-form $G$:
\begin{align}
G^{\mu\nu}=-2\nabla^{[\mu}k^{\nu]}- k^{[\mu}F^{\nu]\sigma}A_\sigma
-\tfrac{1}{2}\,\left(\varphi-\theta \right)\,F^{\mu\nu}\,.
\end{align}
The equations of motion then imply that
\begin{align}\label{geeeqn}
\nabla_{\mu}G^{\mu\nu}=&\left(\alpha-6 \right)\,k^{\nu}+\tfrac{1}{2}F^{\nu\rho}s_{\rho}
-\tfrac{1}{2}({\cal L}_kF)^{\nu\rho}A_{\rho} \,,
\end{align}
where ${\cal L}_k$ is the Lie derivative with respect to $k$.
In our case $k=\partial_t$ and at linearised order
$\varphi=a_{t}+\delta a_{t}$, $\theta=-a_{t}-\delta a_{t}$, $s=-E_{i}\,dx^{i}+a_{t}\,\zeta_{i}\,dx^{i}$
and $\alpha\ne 0$ when $\zeta\ne 0$.

We can now define the following bulk current density
\begin{align}
Q^i=\sqrt{-g}G^{ir}\,.
\end{align}
From \eqref{geeeqn} we deduce, in particular, that
\begin{align}\label{Qeqns}
\partial_iQ^i=0\,,\qquad
\partial_rQ^i=-\partial_j\left(2\sqrt{-g}G^{ji}\right)\,.
\end{align}
By calculating the holographic stress tensor, $t^{\mu\nu}$, we find
\begin{align}\label{hcdef}
\bar G^{3/2} \sqrt{\bar g} t^{ti}-\mu J^i|_\infty = Q^i|_\infty- t \bar G^{3/2} \sqrt{\bar g} t^{ij} \zeta_{j} \, ,
\end{align}
and we conclude that $Q^i|_\infty$ is the time independent part of the heat current density\footnote{The time dependent piece in \eqref{hcdef} implies the static susceptibility of the heat current two-point Greens function is proportional to
$t^{ij}$; see 
\cite{Donos:2014cya,Donos:2014yya}.}.

\subsection{Navier-Stokes on the horizon}
The next step in our analysis is to examine the equations of motion for
the perturbed black holes in the context of a Hamiltonian decomposition
with respect to the radial direction. As is well known the Hamiltonian is simply 
a sum of constraints. We want to evaluate the constraints
as an expansion in the radius at the black hole horizon. The details
of this calculation are technically involved and will be fully described in
\cite{Banks:2015wha}. The final results, however, are simple to explain.
We find that the Gauss law constraint implies that $\partial_i J^i_{(0)}=0$ where 
$J^i_{(0)}\equiv\left.J^i\right|_H$. Furthermore, the $t$ components of the momentum
constraint, $H_t=0$, as well as the Hamiltonian constraint, $H=0$, each separately
imply that $\partial_{i} Q^i_{(0)}=0$ with $Q^i_{(0)}\equiv\left.Q^i\right|_H$. 
Finally, the $i$ components of the momentum constraint, $H_i=0$, gives additional
equations which, when combined with the others, gives the linearised Navier-Stokes equations on the black hole horizon, presented below. 

To summarise, evaluating the constraints at the horizon leads to a closed system of equations for a {\it subset} of the perturbation which  
must be satisfied at the black hole horizon. 
The black hole horizon is as in \eqref{eq:DC_ansatz}-\eqref{nhexpbh}. The perturbation at 
the horizon is given as in \eqref{pertone}-\eqref{eq:nh_exp} and it is illuminating to introduce the following
notation:
\begin{align}\label{defveep}
&v_{i}\equiv-\delta g_{it}^{(0)},\qquad w\equiv \delta {a}_{t}^{(0)}\,,\nn
&p\equiv -4 \pi T\frac{\delta g_{rt}^{(0)}}{G^{(0)}}-g^{ij}_{(0)}\nabla_{j}\,\ln G^{(0)}\delta g_{it}^{(0)} \,,
\end{align}
The current densities at the horizon can be written
\begin{align}\label{eq:JQ_hor}
J^i_{(0)} &=\rho_H v^i+\sigma_H^{ij}\left(\partial_j w+E_j\right)\,,\nn
Q^i_{(0)} &=Ts_Hv^i\,,
\end{align}
where we define the horizon quantities
\begin{align}
\rho_H&=\sqrt{g_{(0)}}a_t^{(0)},\qquad s_H=4\pi \sqrt{g_{(0)}}\,,\nn
\sigma^{ij}_H&=\sqrt{g_{(0)}}g^{ij}_{(0)},\qquad \eta_H=\frac{s_H}{4\pi}\,.
\end{align}
The four unknowns in \eqref{defveep}
satisfy the following system of four linear partial differential equations:
\begin{align}
&\nabla_{i} v^{i}=0\,, \label{eq:v_eq2}\\
&\nabla^2w+v^{i}\,\nabla_{i}({a_{t}^{(0)}})=-\nabla_{i} E^{i}\,,\label{eq:w_eq2}\\
&\eta_H\left[-2\,\nabla^{i}\nabla_{\left( i \right. }v_{\left. j\right)}
+ \nabla_{j}\,p\right]=Ts_H\zeta_{j}
+\rho_H\left(E_j+\partial_j w\right)\,,\label{eq:V_neutral2}
\end{align}
where the covariant derivatives are with respect to the metric on the black hole horizon $g^{(0)}_{ij}$ and all indices are being raised and lowered
with this metric. 
The first two equations are simply $\partial_{i} Q^i_{(0)}=\partial_iJ^i_{(0)} =0$.
Note that in \eqref{eq:V_neutral2} we can also write
$2\,\nabla^{i}\nabla_{\left( i \right. }v_{\left. j\right)}=\nabla^2v_j+R_{ji} v^i$.

Remarkably, we have obtained the time-independent, linearised Navier-Stokes equations for 
a forced, incompressible, charged fluid on the curved horizon. Such equations are also called Stokes equations. The fluid velocity is $v_i$, the effective pressure
is $p$ and $w$ is a scalar potential. The forcing terms are given by the one-forms
 $4\pi T\zeta$ and $E$. 
In \eqref{eq:JQ_hor} and \eqref{eq:V_neutral2}, we see that $\rho_H$, $s_H$, $\sigma_H^{ij}$
and $\eta_H$ can be viewed as coefficients in the constitutive relations for the horizon fluid:
$\rho_H$ and $s_H$ are the charge density and the entropy density while
$\sigma_H^{ij}$ and $\eta_H$ are transport coefficients associated with the electric conductivity and
the shear viscosity of the horizon fluid, respectively. We stress that $\eta_H, \sigma_H^{ij}$ are {\it not}, in general, the shear viscosity and DC electric conductivity of the deformed dual CFT\footnote{Some examples where the horizon transport quantities are, accidentally, the same as the in the field theory case are discussed in \cite{Iqbal:2008by},\cite{Policastro:2001yc,Kovtun:2004de}.}.
It is also worth noting that possible thermoelectric transport coefficients
for the black hole horizon fluid $\alpha_H,\bar\alpha_H$ and $\bar\kappa_H$ are all absent in the expressions \eqref{eq:JQ_hor}.

We now establish a number of interesting properties of this set of equations.
Firstly, we multiply \eqref{eq:V_neutral2} by $v_{j}$ and then integrate over the horizon,  
leading to
\begin{align}\label{positiveexp}
&\int d^{2}x\, \sqrt{g_{{0}}} \left[2\nabla^{\left( i \right. }v^{\left. j\right)} \nabla_{\left( i \right. }v_{\left. j\right)}+\left(\nabla w+E \right)^{2}
\right]\nn
&\qquad\qquad\qquad\qquad=
\int d^{2}x\left( Q^i_{(0)}\zeta_i+ J^i_{(0)} E_i\right)\,.
\end{align}
In the case of non-compact horizons we have assumed that possible boundary terms vanish.
Observe that the left hand side is a manifestly positive quantity and this is associated 
with the thermoelectric conductivities being a positive semi-definite matrix.

Second, we consider the issue of uniqueness for \eqref{eq:v_eq2}-\eqref{eq:V_neutral2}. 
If we have two solutions then the difference of the functions will
satisfy the same equations but with vanishing forcing terms, $\zeta=E=0$. 
Denoting the difference by $(v_i,w,p)$, we immediately conclude from \eqref{positiveexp}
that
\begin{align}
\nabla^{\left( i \right. }v^{\left. j\right)} =0\,,\quad
\nabla_iw=0\,.
%\quad v^i\nabla_i\phi^{(0)}=0\,.
\end{align}
We also have $v^i\partial_i a_{t}^{(0)}=0$ from \eqref{eq:w_eq2} and $\nabla p=0$ from \eqref{eq:V_neutral2}.
We conclude that the solution space is unique up to Killing vectors of the horizon metric, with $p,w$ constant and $\delta g_{rt}$ fixed by \eqref{defveep}.
This results agrees with the intuition that one should be able to boost along the orbits
of Killing vectors to obtain a solution with momentum.

Third, we observe that when $(E,\zeta)$ are exact forms, $(E,\zeta)=(de,dz)$ with $e,z$ globally defined
functions on $\Sigma_2$, we can solve equations \eqref{eq:v_eq2}-\eqref{eq:V_neutral2} by taking
$w=-e$ and $p=4\pi T z$, plus possible constants, and $v^i=0$. We observe that this solution gives zero 
contribution to the current densities \eqref{eq:JQ_hor}
at the horizon. This solution gives no
contribution to the DC thermoelectric conductivity, which we discuss below. Thus, the DC conductivity is determined by the harmonic part of  
$E$ and $\zeta$.

\subsection{The Thermoelectric DC conductivity}
We have shown that the electric and heat currents at the horizon, given in
\eqref{eq:JQ_hor}, can be expressed in terms of the 
sources $E,\zeta$ after solving 
the Stokes
equations \eqref{eq:v_eq2}-\eqref{eq:V_neutral2}. 
To obtain DC conductivities of the field theory we need to relate the currents at the black hole horizon
to the currents at the AdS boundary. In some cases these are the same. 
In general, however, the currents depend on the radial coordinate $r$ 
and one needs to integrate \eqref{jandfeoms}, \eqref{Qeqns}.
In general, however, we can always define total current fluxes
which are independent of $r$
and hence obtain associated DC conductivities.

As a concrete example, assume\footnote{Other topologies will be discussed in \cite{Banks:2015wha}.} we have a periodic holographic lattice on $\mathbb{R}^{1,2}$. That is, the lattice deformations $\bar G(x), \bar g_{ij}(x)$ and
$\mu(x)$ in \eqref{GandF} are periodic functions of the spatial coordinates $(x^1,x^2)\sim (x^1+L_1,x^2+ L_2)$. 
Defining the total electric current flux densities through the $x_2$ plane or the $x_1$ plane, respectively,
\begin{align}\label{avecur}
\bar J^1\equiv \frac{1}{L_2}\int J^1 dx^2\,,\qquad
\bar J^2\equiv \frac{1}{L_1}\int J^2 dx^1\,,
\end{align}
and defining $\bar Q^i$ in a similar way,
we can immediately deduce from \eqref{jandfeoms},\eqref{Qeqns}
that $\partial_r\bar J^i=\partial_r\bar Q^i=0$, which is just Stokes' theorem in the bulk.
These current fluxes at the AdS boundary are thus given by their values at the black hole horizon
which in turn are fixed by $E,\zeta$ after solving \eqref{eq:v_eq2}-\eqref{eq:V_neutral2} to
in order to obtain $v^{i}$ and $w$ and then using \eqref{eq:JQ_hor}. 
This data then gives
the DC thermoelectric conductivities via \eqref{bigform}.

\section{Examples}
For the special case that the lattice depends on only one of the spatial coordinates we can explicitly solve the Stokes equations and recover the results of \cite{Donos:2014yya}. We will present the details of this calculation in \cite{Banks:2015wha}.
Here we will consider the case of a perturbative and periodic lattice
associated with coherent metals with Drude peaks\footnote{the one-dimensional case was studied in \cite{Hartnoll:2012rj,Blake:2013owa}}. 
Specifically, we consider perturbative solutions about
the AdS-RN black brane with a flat horizon. If $\lambda$ is the expansion parameter, at the horizon we assume
\begin{align}
G^{(0)}&=f_{(0)}+\lambda\,f_{(1)}+\cdots\,,\quad
g^{(0)}_{ij}=g\,\delta_{ij}+\lambda\,h^{(1)}_{ij}+\cdots\,,\nn
{a_{t}^{(0)}}&=a+\lambda\, a_{(1)}+\cdots\,,
\end{align}
with $f_{(0)}$, $g$ and $a$ constants, and the remaining functions are periodic on the torus $\Sigma_2$.
For the Ricci tensor we have
$R_{(0)}{}_{ij}=\lambda\,R^{(1)}_{ij}+\lambda^{2}\,R^{(2)}_{ij}+\cdots$ and similarly for the Christoffel symbols.

We can solve \eqref{eq:v_eq2}-\eqref{eq:V_neutral2} perturbatively,
using the expansion: $v=\frac{1}{\lambda^{2}}\,v_{(0)}+\frac{1}{\lambda}\,v_{(1)}+\cdots$,
$w=\frac{1}{\lambda}\,w_{(1)}+w_{(2)}+\cdots$ and $p=\frac{1}{\lambda}\,p_{(1)}+p_{(2)}+\cdots$.
At leading order we find that
$\partial_{i}\,v_{(0)}^{i}=0$, $\Box\,v^{i}_{(0)}=0$
implying that $v^{i}_{(0)}$ are constant on the torus. At next order we have
$v_{(1)}^{i}=N^{i}_{(1)}{}_{j}\,v^{j}_{(0)}$ with
\begin{align}
N^{i}_{(1)}{}_{j}&=-\Box^{-1}\,\left(\partial^{k}(\Gamma^{(1)}){}^{i}_{kj}+R^{(1)i}{}_{j} - \partial^{i}\,\left( \Box^{-1}\partial_{j}R^{(1)}\right)\right).\label{eq:v1_sol_2}
\end{align}
Note that the function $\Box^{-1}f$ is defined up to a constant on a torus. Such constants are fixed at third order in the expansion and they do not affect the DC result at leading order.

We next integrate equation \eqref{eq:V_neutral2} on the black hole horizon, discarding any boundary terms in the non-compact case, 
and keep the $\lambda^0$ part. 
After defining $\Sigma_j\equiv\left(4\pi T\,\zeta_{j}+a\,E_{j}\right)$, now with 
$\zeta_i,E_i$ constant on the torus, we deduce that at leading order in $\lambda$ we have
\begin{align}
v^{i} \approx (L^{-1})^{ij}\Sigma_j,\quad
J^{i}_{(0)}\approx {\rho_H}v^i,\quad
Q^{i}_{(0)}\approx {Ts}v^i\,.
\end{align}
Here $\rho_H=g^2a$ is the charge density at the horizon, $s=4\pi g^2$ is the entropy density, and the constant matrix $L$ is given by
\begin{align}
&L_{ji}=\lambda^{2}g^{-1}\int_H\Big( \frac{g^{-1}}{2} \partial_{j}h^{(1)}_{kl}\partial_{i}h^{(1)}{}^{kl}+\partial_{j}h^{(1)}_{kl}\,\partial^{k}N^{l}_{(1)}{}_{i}\nn
&+\frac{1}{2}\,h^{(1)}\,\partial_{j}(\Box^{-1}\partial_{i}R^{(1)})
+  g^2a_{(1)}\,\partial_{j}\left( \Box^{-1} \partial_{i}a_{(1)}\right)\Big)\,,
\end{align}
where $N_{(1)}$ is given by \eqref{eq:v1_sol_2} and $h^{(1)}=h^{(1)}_{ij}\delta^{ij}$.

Using \eqref{avecur} we can write $\bar J^{i}\approx {\rho}v^i$,
$\bar Q^{i}\approx {Ts}v^i$ where $\rho=(L_1L_2)^{-1}\int dx^1dx^2\sqrt{-g}F^{tr}|_\infty=\rho_H$ is the total averaged charge density. Hence, from \eqref{bigform}
the DC conductivities are given by: $\bar\kappa^{ij}=(L^{-1})^{ij}4\pi sT $,
$\alpha=\bar\alpha=(L^{-1})^{ij}4\pi \rho $,
$\sigma=(L^{-1})^{ij}4\pi \rho^2/s$. 

It is interesting to observe for this general class of holographic lattices that at leading order $\bar\kappa^{ij}(\sigma T)^{-1}_{jk}=s^2/\rho^2\delta^i_k$, corresponding to a kind of Wiedemann-Franz law. Also the thermal conductivity when $J=0$, $\kappa\equiv \bar\kappa -T\bar\alpha\sigma^{-1}\alpha$, as well as the electric conductivity when $Q=0$, $\sigma_{Q=0}\equiv \sigma -T\alpha\bar\kappa^{-1}\bar\alpha$
appear at order $\lambda^0$ in the expansion.
These general results complement those using other techniques in \cite{Mahajan:2013cja}.

\subsection{Acknowledgements}
The work is supported by STFC grant ST/J0003533/1, EPSRC grant EP/K034456/1,
 and by the European Research Council under the European Union's Seventh Framework Programme (FP7/2007-2013), ERC Grant agreement ADG 339140.

\end{document}